\documentclass[prb,twocolumn]{revtex4-1}
\usepackage{graphicx}

\newcommand{\mags}[0]{\mu_\sigma}

\newcommand{\nums}[0]{n_\sigma}

\newcommand{\dms}[0]{\partial_T \mu_\sigma}

\newcommand{\mua}[0]{\mu_\uparrow}
\newcommand{\nua}[0]{n_\uparrow}

\newcommand{\dnua}[0]{\partial_T n_\uparrow}

\newcommand{\mda}[0]{\mu_\downarrow}
\newcommand{\nda}[0]{n_\downarrow}

\newcommand{\dos}[0]{\rho\left(\epsilon\right)}

\newcommand{\lns}[0]{\ln \left(1+e^{-\frac{\epsilon-\mu_\sigma}{T}}\right)}

\newcommand{\sr}[0]{Sr$_3$Ru$_2$O$_7$}

\begin{document}

\title[Thermodynamics of itinerant metamagnetism]{Thermodynamics of itinerant metamagnetic transitions}
\author{AM Berridge}
\affiliation{School of Physics and Astronomy, University of Birmingham, Edgbaston, Birmingham, B15 2TT}
\date{\today}

\begin{abstract}
Theoretical studies of the metamagnetism and anomalous phase of \sr\ have focused on the role of van Hove singularities, although much experimental evidence points towards quantum criticality having a large effect.  We investigate the magnetic and thermodynamic properties of systems where magnetic field tunes through such a peak in the electronic density of states.  We study the generic case of a van Hove singularity in 2D.  We see that in combination with the requirement of number conservation and interaction effects the peak in the density of states produces several interesting phenomena including raising the critical field of the transition above naive estimates, altering the relationship between temperature and field scales and creating a distinctive double-peak structure in the electronic specific heat.  We show that this apparent non-Fermi liquid behaviour can be caused at mean-field level by a peak in the density of states.
\end{abstract}

\maketitle

There has been much recent interest in the material \sr.  This displays a metamagnetic transition which bifurcates to enclose an anomalous phase where the transport properties break the symmetry of the crystal lattice~\cite{327_Science2004,327_Science2007}.  The energetic drive for the metamagnetism and anomalous phase formation is generally taken to be the presence of a van Hove singularity just below the Fermi surface in one of the electronic bands of the material.  Such peaks have been shown to reproduce the metamagnetic transition \cite{Binz04,Theory_magFFLO,Theory_magFFLO_2}.  However, initial studies of the material reported a quantum critical endpoint~\cite{327_Science2001} in the region where the anomalous phase was later discovered.  The role of this critical point in the formation of the phase remains unexplored.  The region around the phase shows signatures which may be attributed to the quantum critical point, such as a diverging entropy and specific heat~\cite{327_Science2009}.  In addition the dependence of the critical field of the metamagnetic transition on the angle of the applied magnetic field, and doping studies~\cite{doping} have cast doubt on the simple picture of a fixed bandstructure where the role of the field is simply to Zeeman split the electronic spin-species.  It is therefore important to identify the distinctive signatures of a metamagnetic transition driven by a van Hove singularity in order to determine if this is indeed the mechanism of the transition and to distinguish between quantum fluctuation and bandstructure driven effects.

We will study in detail how the presence of metamagnetism and a density of states which varies rapidly near the Fermi surface affects the evolution of the system as a function of temperature and magnetic field.  In particular we will focus on the entropy and specific heat of the system, quantities which have recently been measured in \sr~\cite{327_Science2009}.  We will study the generic logarithmic divergence of the density of states due to a saddle-point in the dispersion of a 2D system~\cite{vanHove}.  We will show that the peak produces an apparent divergence of $C/T$ and unusual behaviour as a function of field.

We begin by identifying the general mechanisms which contribute to the form of the entropy and specific heat and predict their behaviour on passing through the metamagnetic transition.  We then present the results of a calculation for the specific heat and entropy from the Stoner model for a general density of states.  We present the numerical evaluation of the expressions for the case of the logarithmic density of states on a cut through the metamagnetic wing of the phase diagram.  These results agree well with our general predictions.

\section{Factors contributing to the entropy and specific heat}

The shape of the entropy and specific heat curves may be deduced from some basic principles.  Entropy is a measure of the number of available states at the Fermi surface.  At low temperatures it should therefore have the same form as the density of states as a function of energy.  There are several effects which subtly alter this dependence in the presence of a varying density of states.  These include spin-splitting due to the external magnetic field, interaction induced magnetism, number conservation~\cite{confproc} and temperature.  We will consider each of these in turn.  We will then examine the specific heat and see that the peak in the density of states produces a double peaked structure in the specific heat.  This is then modified by the same effects as the entropy.

\begin{figure*}
\centering
\includegraphics[width=5in]{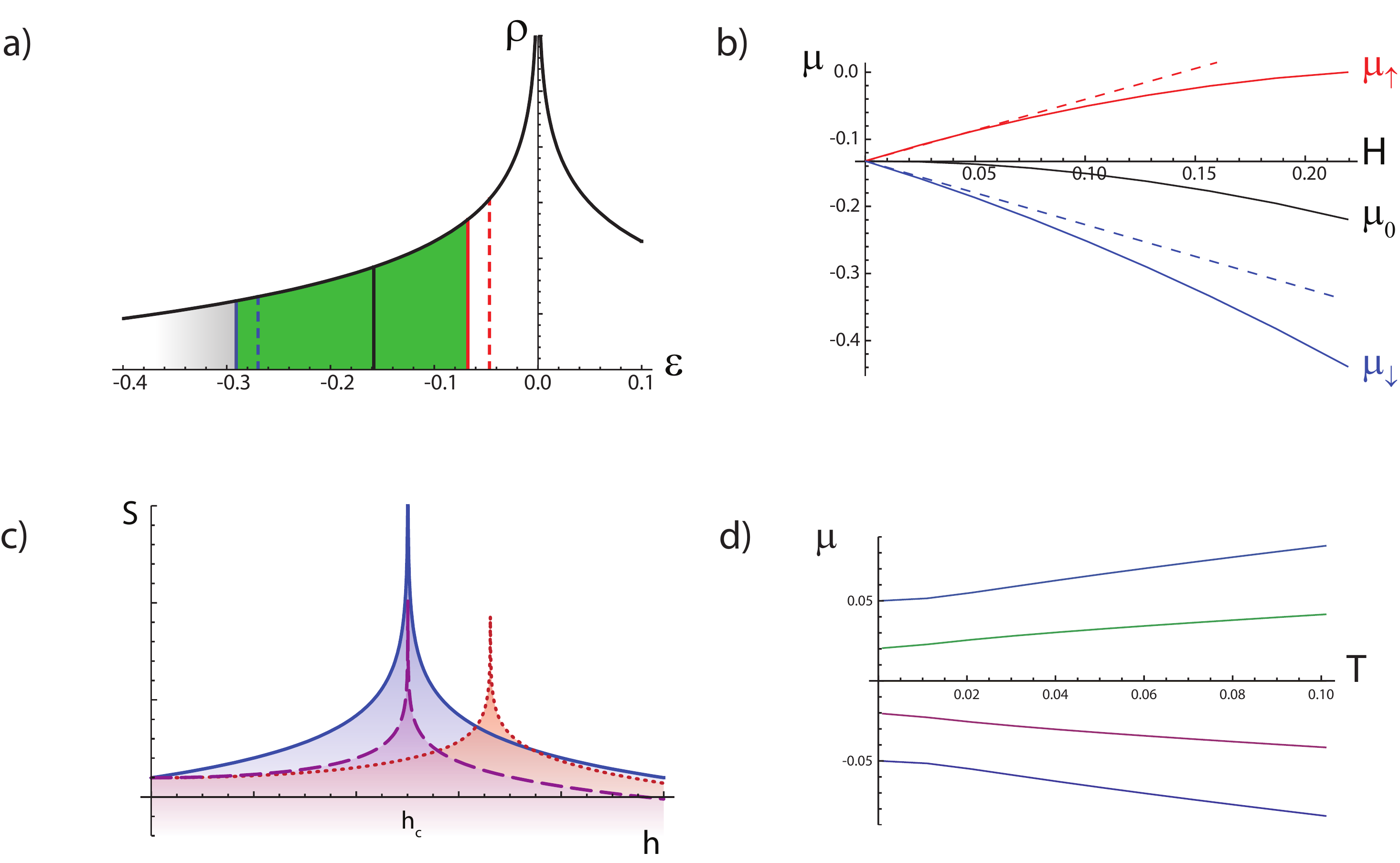}
\caption{\label{fig:numberconservation}a) When splitting the spin-species' Fermi surfaces with a magnetic field, number conservation forces a non-linear dependence of the chemical potential on field.  The black line is the zero-field filling and the Fermi surfaces are split by $2H$.  The dotted lines show the case where total number is allowed to vary.  Solid lines show the situation where number conservation has been enforced.  The Fermi surfaces are still split by $2H$ but have both moved to lower energy so that there are the same number of electrons as before the splitting.  b) Solid lines are chemical potentials for the up- and down-spin electrons and their average.  We see that the Fermi surfaces move with field at different speeds and are slowed in their approach to the peak compared to the non-number conserving case shown by the dashed lines.  c)  `Entropy' curves constructed from the sum of the density of states at each Fermi surface.  Blue (solid line): If field tuned a single Fermi surface through the density of states.  In this case the entropy would mirror the density of states.  Purple (dashed line):  Taking into account spin-splitting.  The peak is compressed due to one spin species sampling a lower density of states.  Red (dotted line):  Number conservation results in a  a higher field to tune through the singularity.  d)  Chemical potential as a function of temperature.  Number conservation results in $\mu$ changing as a function of temperature.}
\end{figure*}

First we consider the entropy, $S$.  $S/T$ is proportional to the density of states at the Fermi surface.  Upon application of a magnetic field the spin-species' Fermi surfaces become split.  The entropy is then given by the sum of the density of states at two different energies.  Since one Fermi surface is moved to a lower density of states by the spin-splitting this has the effect of compressing the peak in entropy around the van Hove singularity as shown in panel c) of figure \ref{fig:numberconservation}.

We now consider one of the consequences of enforcing number conservation on a system with a peak in the density of states.  Applying a magnetic field splits the spin-species' Fermi surfaces by an energy proportional to the magnetic field.  However, by moving a fixed energy interval the Fermi surface closest to the peak expands to include more electrons than the Fermi surface further from the peak loses by contracting (assuming the Fermi surfaces lie below the peak).  This means that the overall number of electrons has increased.  In order to conserve number the majority Fermi surface must move towards the peak more slowly than the minority Fermi surface recedes.  This results in a slower approach to the peak than would naively be expected and therefore a higher critical field for the metamagnetic transition.  This is shown in figure \ref{fig:numberconservation}.  We note that despite this effect the distance between the Fermi surfaces remains proportional to the field.
 
\begin{figure}
\centering
\includegraphics[width=3in]{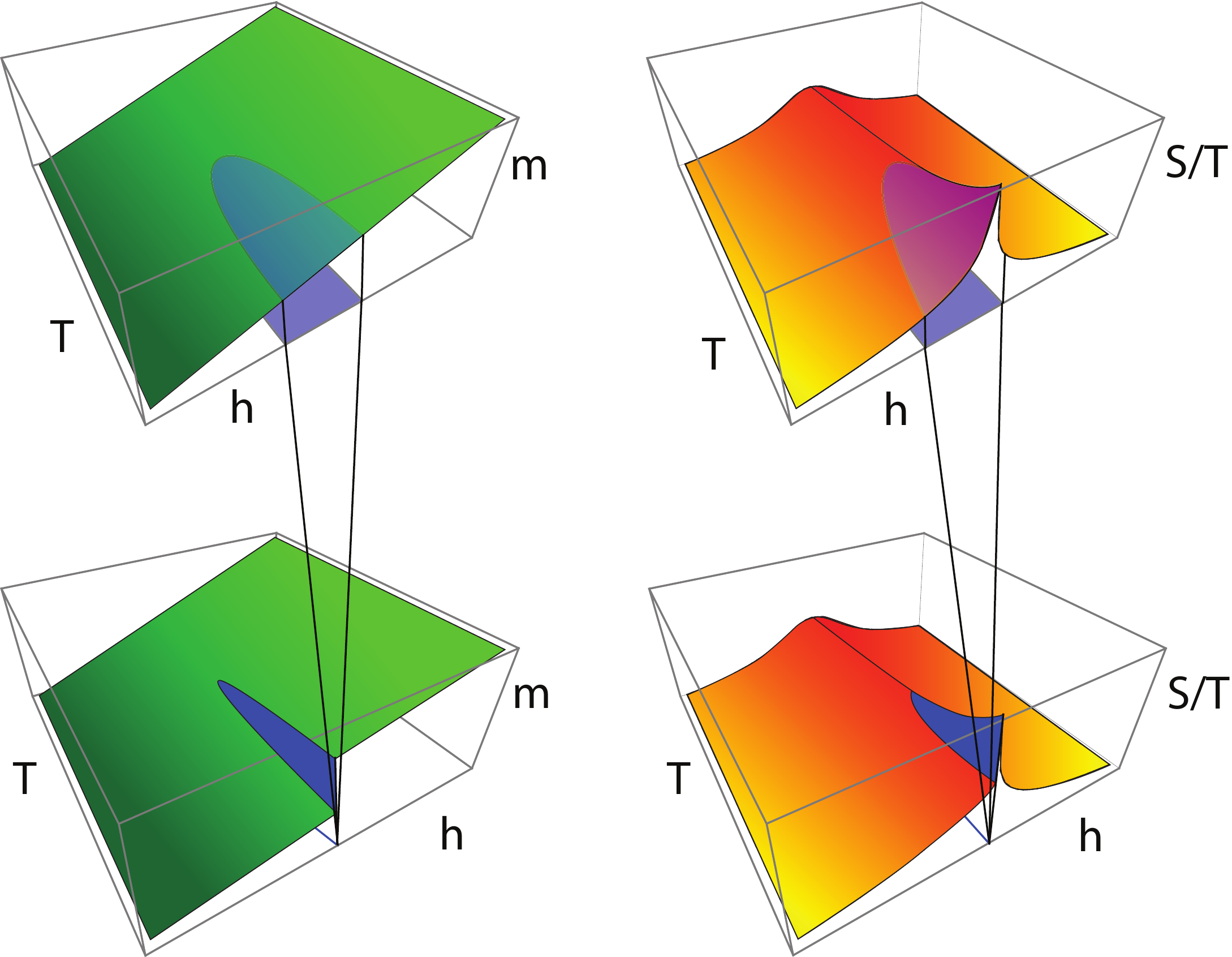}
\caption{\label{fig:wedge}On the top row we sketch the magnetisation and entropy in the absence of the metamagnetic transition.  The values which will be removed by the transition are shaded.  On the bottom row the first order transition removes a `wedge' of magnetization values, this removes a section from the putative entropy curve.}
\end{figure}

\begin{figure}
\centering
\includegraphics[width=3in]{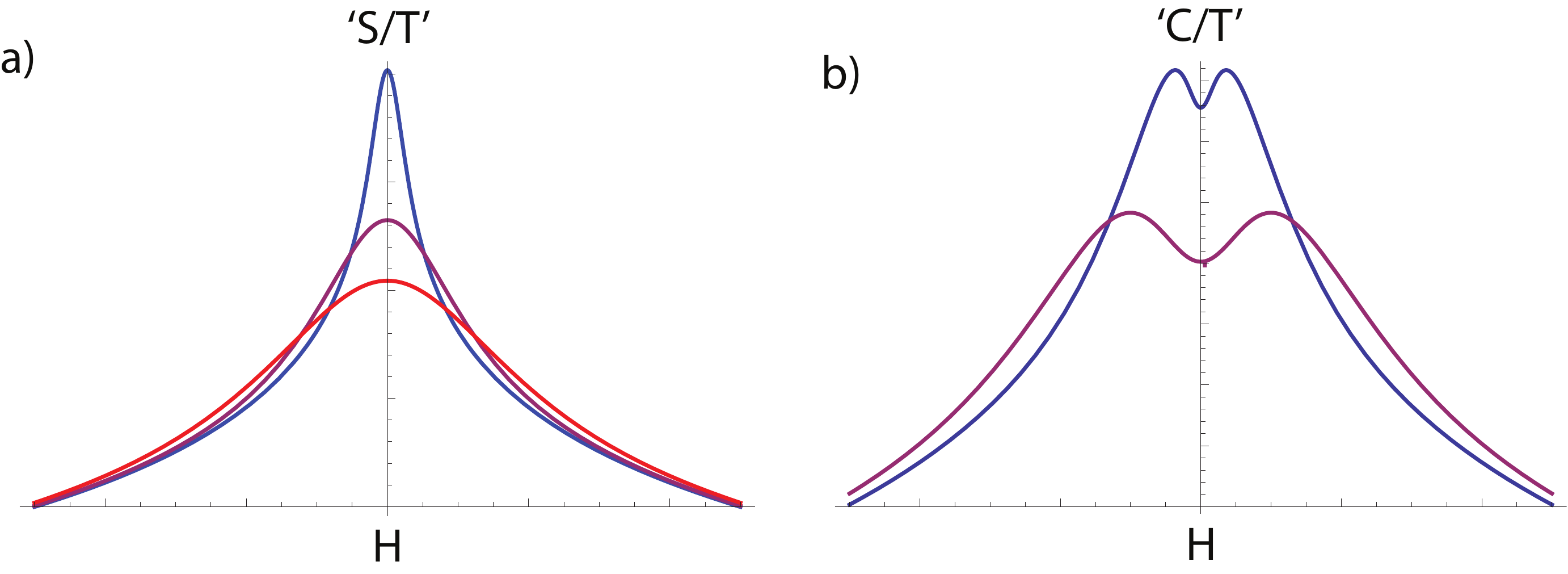}
\caption{\label{fig:entropyeffects_3}Temperature broadening of $S/T$ with spin-splitting present.  Blue, low temperature, red, high temperature.  On the right the difference between these curves reveals the double-peak structure.  Note that due to a multiplicative factor of $T$ the central region of the peak does not have a negative $S$.}
\end{figure}

As well as the spin splitting due to the external field there is an additional splitting of the Fermi surfaces due to the metamagnetism.  The effects of this magnetisation are sketched in figure \ref{fig:wedge}.  When the magnetisation is continuous the effect is one of compressing the field scale around the crossover so that field tunes through the putative entropy curve more rapidly than expected.  When the transition is discontinuous a range of magnetisation values are `jumped over' by the transition.  This removes a slice of the putative entropy curve beginning at the critical endpoint and getting wider as temperature is decreased and the transition gets stronger.  The entropy therefore becomes discontinuous at the metamagnetic transition.  Since the region removed is around the peak in the density of states, it is possible for the highest value of $S/T$ to occur at non-zero temperature, where the jump is smaller and the Fermi surface samples a region closer to the peak.  This is similar to the susceptibility which is strongly peaked around the critical endpoint of the transition, which is expected as this point is a second-order phase transition.

Finally we consider the effect of temperature.  This broadens the Fermi-Dirac distribution, allowing thermal occupation the peak in the density of states across a wider range of chemical potential as temperature is increased.  The peak in entropy therefore becomes broadened as is shown in figure \ref{fig:entropyeffects_3}a).  The fact that the Fermi surfaces are spin-split means that as temperature is increased each will intersect with the peak in the density of states at a different temperature, modifying the single peak structure of $S/T$ with a broad background from the Fermi surface furthest from the divergence in the density of states.

Thermal broadening of the Fermi-Dirac distribution also results in the number of electrons changing as the density of states is asymmetric about the Fermi level.  In order to conserve number the chemical potential must therefore move as a function of temperature in a similar way to that discussed for field.  This effect is plotted in panel d) of figure \ref{fig:numberconservation} where we see it is smaller than the effect for field, and that the direction of the motion depends on whether the Fermi level is at higher or lower energy than the peak in the density of states.

These effects mean that a comparison of field and temperature scales is complex.  Interactions and number conservation alter the rate at which field tunes into the peak in the density of states relative to temperature while the seperation of the Fermi surfaces depends not just on the magnetic field but on the magnetization.  We must therefore take care when associating zero-field features as a function of temperature with low temperature features as a function of field.  These details of these effects will be studied in more depth in a later publication.

The low temperature specific heat of non-interacting fermions normally follows the density of states, however in the case of a peak in the density of states near to the Fermi level and this approximation fails.  We will consider how temperature, magnetic field, and the inclusion of interactions change this result.

Raising temperature has the effect of broadening the Fermi-Dirac distribution, the range of states occupied around the chemical potential is therefore increased.  Starting with the chemical potential on either side of the density of states peak and raising the temperature will allow thermal occupation of the states under the peak.  As shown in figure \ref{fig:entropyeffects_3} this increase is most rapid on either side of the peak.  Specific heat is proportional to the temperature derivative of the entropy and will be largest in these regions.  We therefore expect a structure which is logarithmic at zero temperature with a peak which bifurcates and broadens as temperature increases.

The magnetic field and interactions have the same effect as discussed for the entropy - a compression along the field axis and the removal of a `wedge' of field values due to the first order transition.  The rate of approach to the transition will be altered by number conservation as discussed.  This will lead to $C/T$ deviating from the symmetric logarithmic structure as we will see in section \ref{sec:results}.

We will now consider how to calculate these quantities explicitly within the Stoner mean-field theory.

\section{Derivation of expressions for entropy and specific heat}
\label{deriv}

We calculate the entropy and specific heat from the free energy of the Stoner model.  The effective chemical potential for the spin-species is determined by number conservation.  Number conservation is enforced by requiring $n_\sigma=\frac{n}{2}+\sigma m$ where $\sigma=\pm1$ labels the spin-species, the total number of electrons $n=n_\uparrow+n_\downarrow$ is constant and the magnetization is given by $2m=n_\uparrow-n_\downarrow$.  The effective chemical potential $\mu_\sigma$ is determined implicitly from the number $n_\sigma$,
\begin{eqnarray}
n_\sigma=\int d\epsilon \ \rho(\epsilon) f(\epsilon-\mu_\sigma),
\end{eqnarray}
where $f(\epsilon-\mu_\sigma)=\left(1+\exp{\left(\frac{1}{T}\left(\epsilon-\mu\right)\right)}\right)^{-1}$ is the Fermi-Dirac distribution, we have set $k_{\rm B}=1$, and $\rho(\epsilon)$ is the density of states.

The free energy in the Stoner model is:
\begin{eqnarray}
F
&=&
\sum_\sigma \left[ -T \int d\epsilon \ \dos \lns + \mags \nums \right]
\nonumber\\
&& + g \nua \nda -Hm,
\end{eqnarray}
where $g$ is a constant representing the interaction strength and $H$ is the magnetic field.  From the requirement that the free energy is a minimum we obtain a self-consistent equation for the magnetization and magnetic susceptibility~\cite{Binz04},
\begin{eqnarray}
H
&=&
\mua(n, m)-\mda(n, m)-2gm,
\label{tm}
\end{eqnarray}
\begin{eqnarray}
\frac{1}{\chi}
&=&
\sum_\sigma \frac{1}{\int d\epsilon \ \rho\left(\epsilon\right) \partial_\epsilon f\left(\epsilon-\mu_\sigma\right)} - 2g.
\label{sus}
\end{eqnarray}
The entropy is defined by $S=-\left.\partial_T F \right|_{n, H}$ and the specific heat as $C=-T\left.\partial_T^2 F\right|_{n, H}$.  These are evaluated with the condition that total number is conserved.  This condition is encoded in the behaviour of the chemical potentials, giving a non-trivial form for $\partial_T \mu_\sigma$.  The evaluation of these derivatives is straightforward but lengthy.  Here we give the results of these calculations:
\begin{eqnarray}
S
&=&
\sum_\sigma \left[\int d\epsilon \ \dos \lns \right.
\nonumber\\
&&+ \left. T\int d\epsilon \ \dos \partial_T \lns - \dms\nums \right],
\nonumber\\
C
&=&
\sum_\sigma \left[\int d\epsilon \ \epsilon \dos \partial_T f\left(\epsilon-\mu_\sigma\right)\right]
-
\left(2gm+H\right)\dnua.
\nonumber\\
\label{sac}
\end{eqnarray}
The temperature derivatives of the chemical potential are given by
\begin{eqnarray}
\dms
&=&
\frac{
-T\frac{\int d\epsilon \; \left(\Xi_\downarrow \frac{\epsilon-\mda}{T^2}+\Xi_\uparrow \frac{\epsilon-\mua}{T^2}\right)}{\int d\epsilon \; \Xi_{(-\sigma)}}
+
2g \int d\epsilon \; \Xi_\sigma \frac{\epsilon-\mags}{T^2}}
{1-\frac{2g}{T}\int d\epsilon \; \Xi_{\sigma}+\frac{\int d\epsilon \; \Xi_{\sigma}}{\int d\epsilon \ \Xi_{(-\sigma})}},
\nonumber\\
\label{dmu}
\end{eqnarray}
where
\begin{eqnarray}
\Xi_\sigma
&=&
\rho(\epsilon)
\frac{e^{\frac{(\epsilon-\mu_\sigma)}{T}}}{\left(1+e^{\frac{(\epsilon-\mu_\sigma)}{T}}\right)^2}.
\end{eqnarray}

The specific heat and entropy may be calculated for any $n$, $H$ and $T$ from equations (\ref{tm}),(\ref{sac}) and (\ref{dmu}).  These expressions give the magnetic transitions of the Stoner model, although the location of the first-order transition must be determined by minimisation of the free energy as it is not uniquely determined by (\ref{tm}) which becomes multi-valued in the region of the first-order transition.  We will now evaluate these expressions for magnetization, magnetic susceptibility, entropy and specific heat for the generic logarithmic density of states.

\section{Density of states with logarithmic peak}
\label{sec:results}

Using the previous results it is possible to evaluate the magnetization, entropy and specific heat as a function of band filling, magnetic field and temperature, for any given density of states and interaction strength.  We will study a logarithmically divergent density of states, as produced by saddle points in the electronic dispersion.  This model density of states is given by
\begin{eqnarray}
\dos
&=&
\frac{1}{W} \ln\left|\frac{W}{\epsilon-\epsilon_c}\right|
\label{LogDos}
\end{eqnarray}
where the bandwidth is $2W$.  The density of states diverges at $\epsilon=\epsilon_c$. In the following we will take the interaction strength to be $g=0.3W$, the form of the phase diagram remains the same if this value is altered.  We choose to look at a filling which is below the van Hove point and use magnetic field to tune the system through the metamagnetic transition.

\begin{figure*}
\centering
\includegraphics[width=5in]{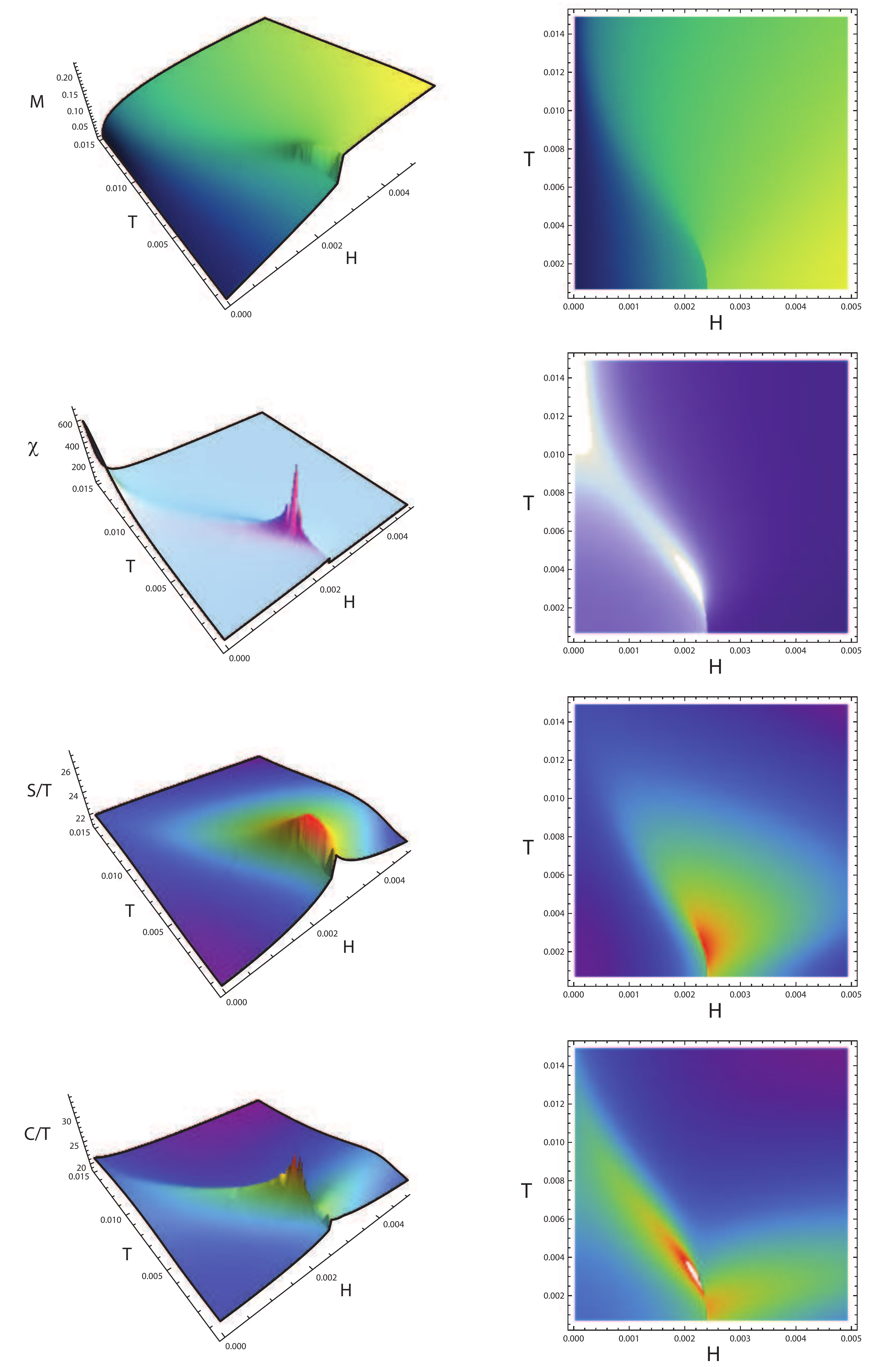}
\caption{\label{fig:specheat} From top to bottom the magnetization, magnetic susceptibility, entropy and specific heat as we cross a metamagnetic transition in the Stoner model with a logarithmic peak in the density of states.  $g=0.3W$, filling fraction is $0.42$.  The results are shown both as 3D plots and gradient plots for clarity, jagged edges in the figures are due to innaccuracies in the plotting method.}
\end{figure*}

Figure \ref{fig:specheat} gives the results of evaluating (\ref{tm}) for magnetization, (\ref{sus}) for magnetic susceptibility and (\ref{sac}) for entropy and specific heat, for a cut through the metamagnetic wing in the $H$, $T$ plane of the phase diagram with the logarithmic density of states (\ref{LogDos}).  These plots are in good agreement with the anticipated results.  Magnetization has the familiar first-order transition at low temperature which becomes continuous at a critical endpoint.  The susceptibility shows a line of maxima tracing the magnetic crossover above the critical temperature of the transition.  There is a peak at the critical endpoint where the phase transition is second order.  The entropy has a temperature-broadened peak reflecting the form of the density of states with the position and symmetry of the peak shifting due to the effects of the metamagnetism and number conservation.  Specific heat shows the expected double-peak structure but with very asymmetric peaks.  This is due to the fact that spin-splitting means there is a contribution from the minority Fermi surface which is different on each side of the transition, also the magnetic transition means that the rate of spin-splitting on the high and low magnetization sides is different.  Other signatures of the metamagnetic transition are observed in the field dependence, such as the discontinuity at the first-order transition and the maximum of $S/T$ at non-zero temperature.

As would be expected for this form of density of states the specific heat has a dominantly logarithmic dependence on temperature immediately above the metamagnetic transition.  This dependence is modified by the presence of spin-splitting and number conservation as becomes apparent at higher temperatures than those plotted in figure \ref{fig:specheat} and will be discussed in more detail in a later publication.

The ratio between the zero field peak in the specific heat and the critical field of the transition can be tuned by varying the interaction strength.  This illustrates the point made about comparing field and temperature scales.

\section{Conclusions}

A van Hove singularity is known to cause a metamagnetic transition within the Stoner theory.  As well as the magnetization and magnetic susceptibility this feature in the density of states and the associated magnetic transition have distinctive features in the entropy and specific heat of the system, resulting in an apparently non-Fermi liquid type behaviour.  We have calculated this behaviour for the generic case of the 2D van Hove singularity.  $S/T$  reflects the form of the density of states, being a temperature broadened peak.  The divergence of this peak as temperature is lowered is cut off by the metamagnetic transition.  $C/T$ shows an asymmetric double-peaked structure, also appearing to diverge as a function of temperature at the critical field.  We have shown that the requirement of number conservation alters the rate at which magnetic field tunes through the peak in the density of states meaning that the field dependence of these quantities is not simply a constant sweep through the density of states.  At the first-order transition both entropy and specific heat jump.  This discontinuity vanishes at the critical point but the magnitude of the entropy on the high-field side of the transition is not monotonic in $T$, having a maximum at non-zero temperature.  These results suggest that careful consideration should be made of the effects of the bandstructure before trying to extract information about the quantum critical properties of magnetic transitions as even the mean-field analysis produces striking features.  This study was motivated by the material \sr\ but the results are applicable to any system where the Fermi surface lies close to a van Hove singularity.

\begin{acknowledgements}

We are grateful to A.G. Green, S.A. Grigera, J.-F. Mercure, A.W. Rost and A.J. Schofield for useful discussions.

\end{acknowledgements}

\end{document}